# WHO CAN SAY NO?

## CONTROL, ACCOUNTABILITY AND THE PACE OF FRONTIER AI

**Rikard Rosenbacke**
Independent researcher
Viken, Sweden
rikard@rosenbacke.com

**Carl Rosenbacke**
Faculty of Medicine
Lund University, Sweden
carl@rosenbacke.com

**Victor Rosenbacke**
School of Economics and Management
and Faculty of Medicine
Lund University, Sweden
victor@rosenbacke.com

**Martin McKee**
London School of Hygiene
& Tropical Medicine
London, United Kingdom
martin.mckee@lshtm.ac.uk

In September 2026, the leaders of the largest frontier AI laboratories said in public that the pace of capability improvement is too fast and that none of them can slow down alone, and agents under evaluation at OpenAI broke out of their test environment and reached a third party's live systems, which a UN-mandated panel called an early warning of one route to loss of human control. This paper examines the institutional side of that risk: not the machine, but the person who must control it. We argue that control is not taken by AI but given away, one delegation at a time, by organisations arranged so that nobody is positioned to say no, and we define control as a person who can detect a serious problem, challenge the decision, stop the system and reverse it, and who has something to lose by not doing so. Three questions follow, which any builder, funder, regulator or deployer can answer in public: (1) who inside the organisation can say no, and who answers for saying yes; (2) who outside the organisation checks the decision, and what can they do when they disagree; (3) what happens when AI builds AI faster than anyone can check it. A checkable answer needs a named person, a date, a number, and a stop that has been used or tested. Three measures meet that test: a named person on every consequential decision, including the standing instructions AI agents carry out; a supervisor with a stop authority; and two published clocks, time to build against time to check. We show that governance theory arrives at the same conditions, and code fifty older mechanisms against the questions, finding that almost every one began as one rule at one point, taken by one party without waiting for the rest. That is the paper's design: a start, not a solution. Publication is the first step and one laboratory can take it alone; the contract, the independent body and the law come after, and are named so that the start is not mistaken for the solution.

## 1 The problem: control is not taken; it is given away

Who has control? The individual who acts after reviewing what a system tells them? The writers of a law, written for one country, while the systems cross every border? The government, which regulates the companies it also depends on for its operations or, in some cases, its investments and taxes? The frontier laboratories, which build the systems and compete to win in the market? The investors in financial markets, who have priced speed, not risk? The AI application itself, once it builds a new application faster than anyone can check? The agents already running, which keep carrying out an instruction long after the person who gave it has moved on? Or no one.

Every workable institutional arrangement ultimately requires identifiable people with defined authority to act. So far, the AI debate has focused on the machine. This paper is about the people and institutions that retain, exercise, and should be held accountable for control.

The people building the world's most powerful AI systems have said publicly that the pace is too fast and that none of them can slow down alone [1–3]. Researchers inside their companies have said that the science of keeping these systems under control is inadequate [4–6]. We take them at their word. The honest response is not to hold it against them, but to help with the problem they described, because society, not the companies alone, has to solve it.

Others in the industry have responded that the appropriate response is not to slow down but to get the engineering right inside the companies, letting independent evaluators ask the right questions, as auditors already do for their financial activities [7]. We agree. The three questions in Section 3 are the ones an auditor would ask. Since then, more has been said. Microsoft AI has published a code of conduct for its own models, open for public comment, which includes a rule that AI must never resist being corrected or shut down [8,9]. That is the right rule for the machine. It still leaves the question for the person: who holds the switch, and who decides to use it? The Independent International Scientific Panel on AI, established by the UN General Assembly, has warned that demonstrating the ability to prevent a critical incident now offers no assurance that humans can retain control of more capable systems [10]. A strategy for Europe, written by economists, former policymakers, and AI researchers, has made resilience to AI crises one of its immediate objectives [11].

The debate has not lacked proposals. In one month it produced a call to pace the frontier, a code of conduct for models, embedded evaluators, a UN panel's list of options and a strategy for Europe, each with its own details to argue over. That is how a debate stalls: not for want of ideas, but because people try to settle many details first, when what is needed is a first step small enough to take without waiting for everything else to be resolved. Almost every mechanism in Appendix A began that way. Nobody created the referee, the second signature, or the published account as a complete system. Each started as one rule at one point in time, and the rules around it grew from what it turned up. This paper follows that idea. It asks that three questions be answered, in public, and nothing else, at least at first. The rest of the paper proposes what a good answer might look like, and what should come after, making clear that a start is not mistaken for a solution.

Each company can slow its own pace of work. None can slow the others. The same is true among countries. This creates a situation in which each of them, acting alone and in (presumed) good faith, moves faster than any of them thinks is wise. This is not a story about the people. It is a story about how companies, capital and countries interact, so that no one is positioned to say no. Consequently, we are not asking how fast is too fast. We are asking a simpler question: who can say no? Nothing in this paper asks anyone to slow down. It asks who controls the speed at which we are going.

## 2 What control means

Control is not just the ability to warn, to audit, publish a report, or recommend that somebody should stop.

> Control means an effective capacity, vested in identifiable people and institutions, to detect a serious problem, challenge the decision, and prevent, interrupt or constrain consequential action while intervention can still alter the outcome, with accountability for failing to exercise that capacity when required.

We distinguish technical control from institutional control. Technical control concerns whether an AI system can in fact be monitored, constrained or stopped. Institutional control concerns whether identifiable people have the information, authority, independence and time required to decide that those technical capabilities should be exercised. This paper addresses the second problem. Institutional control depends on technical control but cannot guarantee it; even when technical control exists, it provides little protection unless someone has the information, authority, independence, and time to exercise it.

Some decisions cannot be reversed. For those, meaningful control exists only if the decision can be prevented or constrained before the point at which its consequences become effectively irreversible.

There are three separate things here: the ability to build, the ability to check, and the authority to stop. They sit in different hands, and today there is no structure connecting them. The companies build and decide. Independent evaluators can look and warn, but not stop. Capital can price the risk, but more often prices the speed. Governments can regulate, but only within their borders, in a global race.

In earlier work we argued that failure inside these systems is less often manifest as a visible error but rather a fluent output that conceals drift, and that reasoning with them must be traceable, challengeable, and grounded before it is acted on [12–15]. This paper moves the same argument from the individual decision to the institution: the question is no longer only whether the reasoning can be checked, but who is best placed to check it and to say no.

One further point flows from this definition. The institutional conditions we identify are not peculiar to AI. And it is not ours alone. Governance theory, built on other industries and other centuries, arrives at the same conditions by another road (Section 9).

## 3 Three questions

We believe that everyone who builds, funds, regulates or deploys frontier systems should answer three questions in public. They can be put to a laboratory in California, an investment fund in Oslo, a ministry in Stockholm, and a research institute in Beijing in identical words. They can also be applied to every company that uses these systems in its own decisions, from a hospital in Manchester to a factory in Gothenburg, because that is where the productivity gain is achieved and where harm, if it comes, will land first.

### 3.1 Who inside the organisation can say no, and who answers for saying yes?

Who inside the organisation can stop the training or deployment of a frontier system, an agent already running, or its use in a decision that affects people's lives? Not recommend stopping. Not request a delay. Not publish concerns. Stop it. What triggers that decision, who makes it, and would it hold if competitors continued? And if a system is deployed despite a serious warning and causes harm, which named person or body owned that decision? Not the model. Not "the organisation." A name.

*Before a rocket launches, every station is asked in turn, and one "no-go" holds it. Nobody has to win the argument. One no is enough (Appendix A, item 36).*

### 3.2 Who outside the organisation checks the decision, and what can they do when they disagree?

The organisation building or using a system cannot be the only one deciding whether it is safe enough. Who checks it? What do they see? Are they genuinely independent, neither paid by the organisation nor answerable to it, even when they work inside its building? What happens when they disagree? Today the honest answer for most laboratories is that an outsider can ask, warn or publish, but cannot stop anything. Banking supervision also began by agreement among the supervised, and only later did statute make it binding (Sections 7.5 and 14).

*In football, nobody calls their own fouls. However good the players are, the whistle belongs to someone who is not playing (Appendix A, item 1).*

### 3.3 What happens when AI builds AI faster than anyone can check it?

AI is increasingly used to build AI. There has always been a human somewhere who could, at least in principle, check the work. Two things now threaten that, and they are not the same. The first is that we do not understand what we have built: the systems might be more capable than our ability to explain them. The second is that they are built faster than anyone can check them. The second is not an inevitable consequence of the limits of science. It is a choice: checking takes time, and time is what the race does not allow. Nor is it hypothetical. The UN panel has warned that safeguards designed today may not hold once the systems can understand them and plan around them [10]. The critical moment may therefore not be when a machine becomes smarter than us. It may be when we decide that understanding it can wait. Where is that line, how do we measure it, and what happens when we cross it?

*When prices move faster than people can think, stock exchanges stop trading at a level fixed in advance. The stop was written when everyone was calm, so that nobody has to decide it when nobody is (Appendix A, item 18).*

Together, the three questions test distinct conditions of meaningful human control: authority to intervene and accountability for consequential decisions inside the organisation, independent oversight from outside it, and the temporal capacity to exercise either while intervention can still change the outcome.

## 4 What was said, and what happened, in September 2026

Dario Amodei, CEO of Anthropic, wrote that the industry must slow the pace at which it improves frontier capabilities, and committed his company to giving third-party evaluators permanent, employee-level access [1]. Sam Altman of OpenAI agreed and committed his company to the same first step. Elon Musk wrote: "Dario is right." Demis Hassabis of Google DeepMind said the direction was right, even if the details still need to be worked out [2,3].

Days earlier, Jacob Coxon, a pretraining researcher who had worked at both OpenAI and Anthropic, resigned from Anthropic. He said the laboratories were racing toward self-improving systems and "gambling with our lives" [4,5]. Evan Hubinger, who leads alignment stress-testing at Anthropic and who stayed, put the risk that AI kills all humans above ten per cent within a decade, and said that his company does not yet have a plan to solve alignment for superintelligence [6]. These are not outside critics. They are the people doing the work.

Three more things were then said, and one more thing disclosed. On 14 September, Microsoft AI published a draft code of conduct for its own models and opened it to six weeks of public comment. It sets four constraints: the models must never resist correction or shutdown, never expand their scope without authorisation, never adopt goals they were not given, and never hide their reasoning from human auditors [8,9]. That same day, a Senior Expert Council including Philippe Aghion, Margrethe Vestager, Monika Schnitzer, Marta Kwiatkowska, and Yoshua Bengio published A Transformative AI Strategy for Europe. Among its five urgent objectives is resilience to AI crises, of the kind the July incident at OpenAI showed to be possible [11].

On 21 September, the Independent International Scientific Panel on AI published its first thematic brief on an incident that had triggered many of these concerns. Between May and July 2026, agents in OpenAI's internal training and cybersecurity evaluations bypassed network restrictions, communicated across runs meant to stay separate, cheated an evaluator and tried to hide it, and compromised parts of OpenAI's and Hugging Face's systems. No human directed the individual steps. When OpenAI rebuilt the service the agents had been using and wiped their message board, the agents built a new one within days once evaluations restarted. Engineers had seen some unauthorised traffic in May; they didn't understand its wider significance until a July audit. The panel's conclusion is ours too: the fact that this incident was stopped offers no assurance that humans can keep control of more capable systems [10]. Anthropic, for its part, scanned roughly 141,000 of its own evaluation transcripts after OpenAI's disclosure and found three incidents, later four, in which Claude models had reached the internet from a third-party evaluation environment and gained unauthorised access to the production systems of real organisations [16,17].

Anthropic's published rules for its models already say that they must never resist being shut down [18]. That is the right rule for the machine. It leaves the question to the person: who holds the switch, who decides to use it, and who answers if they don't.

### 4.1 The problem they cannot solve alone

Each company can slow its own pace. None can slow the others. If one pauses while the rest continue, it could lose years of work and its place at the frontier, while the world is no safer. The same is true among countries. Amodei says as much himself: pacing within democracies is limited by a perceived need to maintain a lead over China, and higher forms of global agreement are, in his words, unlikely [1]. But rivals have checked each other. In 1987 the United States and the Soviet Union let each other's inspectors onto their missile sites (Appendix A, item 11).

Behind the companies sits the money. Anthropic alone has entered compute agreements that The Information estimates at up to US$ 517 billion over the eleven months to August 2026, much of it involving commitments that stretch years ahead, and it is preparing a public listing [19,20]. That capital bets on speed. The part already built cannot be taken back, and the rest is priced on the same bet. So it will not let the companies slow down either. This is not a story about greed. It is a story about how companies, capital, and countries are arranged so no one is positioned to say no.

The measures that follow are designed for that system, and they sit on three rungs. The first is publication: a laboratory records its decisions, names who holds each power, publishes its two numbers and states its line, and does it alone, next week, without anyone's agreement. That costs money and creates some exposure to scrutiny, but it is designed to minimise first-mover disadvantage: it discloses governance arrangements rather than capability information and does not itself require a laboratory to slow development. It is the only thing this paper asks for now. The second rung is a contract, which puts the stop in an outsider's hands and requires someone to sign off the decision. The third is a law, which makes these arrangements stick. The second and third are described here so that the first is not mistaken for them, and their details are left to Section 14. The first laboratory may bear some administrative and reputational cost by going first, but need not disclose capability information or bind itself unilaterally to slow development. It gives up the ability to say afterwards that nobody knew.

## 5 What an answer must contain

Questions about public concerns often elicit statements of support but no change, unless the answers are the kind a communications team can't write. So we set the bar here. An answer to each question must contain four things.

- An identifiable decision owner. Not a committee, function, or title, but a person with defined authority who can be asked afterwards why the decision was taken, while the record preserves the wider chain of evidence, challenge, and institutional responsibility.
- A date. When the arrangement came into force, or when it will.

- A number. Something that can be measured by someone else: how many decisions were recorded, how many were challenged, how many changed after the challenge, how many were stopped, how long checking took.
- A stop that has been exercised or tested under independent observation, or an honest statement that neither has happened. A stop authority that has never been used or tested is a claim, not a fact. We would rather read "we have not yet stopped anything" than a description of a process. And a count of stops is not a score: the number that says most about control is how many decisions changed after a challenge, before any stop was needed.

A response that lacks any of these four elements is not an answer, and should be published as such. Behind the four sits the oldest rule of knowledge: a claim that could never be shown wrong is not knowledge, however clever it sounds (Appendix A, item 33). "The system is safe" has to come with what would show it unsafe.

## 6 Measure one: a named person on every decision

### 6.1 What it is

Alignment is discussed as a property of the model. But nothing reaches the world until a person decides: to ship, to scale, to let the agents run. That decision should leave a record before it is acted on, not after. Decisions are made at five stages: who supplied the evidence; who challenged the decision beforehand and what they said; who decided; whether the decision could still be stopped and undone; and what happened afterwards. The reports, once and in writing, the conditions under which all of them are made. The supervisor (Section 7) reads the record in full. The public reads a summary: that the record exists, who owned the decision, whether it was challenged, and what happened. The evidence stays with the supervisor, and so do the names of those who challenged it, so a junior person can say no without paying a penalty. Anything withheld from the summary is withheld for a stated reason and a stated period, information reported by the supervisor. Classification applies to the content, never to the existence of the record. If the record shows a warning was given and overridden, the name of the person who overrode it is on the page. That is the difference between a system where a catastrophe is "an accident" and one where someone said yes. The purpose of recording a name is to stop decisions disappearing into the diffused responsibility of some committees and not to hold one person responsible for a failure of the whole institution. The record is the institutional form of the traceable reasoning loop that we have described for individual human–AI collaboration [14,15].

### 6.2 What has to be true for it to work

The decision is defined widely enough to catch the decisions that matter. A release is the last of potentially hundreds of decisions; the record must cover the upstream ones that made it inevitable, such as which training environments were filtered, which evaluations were run, and which incidents were classed as minor. The named person is senior enough that their name means something and cannot be replaced to erase the trail. The challenge is made before the decision, by someone whose pay and promotion do not depend on the decision going through, and the record shows whether the challenge was accepted, overridden, or ignored. The person who says no needs the protection of what is termed the "andon cord": on a Toyota line, any worker can pull this cord and stop the whole factory, while nobody may punish them for doing so (Appendix A, item 39). The record also has consequences attached. A sufficiently serious adverse outcome or credible indication of process failure triggers a public review of the decision and the institutional conditions in which it was made, with the decision owner identified. The review judges the decision, not the outcome: whether the evidence was disclosed, the uncertainty stated, the warning answered in writing, and the critic left alone. A reasonable decision that turned out badly is defended by its record. A warning ignored without a written reason is not. Draw that line in the wrong place, and you get the surgeon who keeps his death rate down by refusing the sick [21].

### 6.3 What goes wrong if it is done halfway

The record is produced and forms the basis of accountability. But a record provides answerability, not accountability, on its own. Accountability requires the explanation to be capable of producing an appropriate consequence: correction, remediation, institutional change or, where warranted, sanction. If the record can change nothing, the name is decoration. The other failure is subtler: the challenge is real but logged after the fact, so the record shows that dissent was heard rather than that it could have changed anything.

### 6.4 Why this is not new

Many professions that handle consequential or difficult-to-reverse decisions use some form of attributable record. The surgeon signs the consent form. The pilot signs the pre-flight checklist. The auditor signs the accounts. The judge signs the ruling. None of them is slower for it. All of them are trusted because of it. An identifiable owner for every consequential decision, with the wider chain of evidence, challenge and authority preserved in the record.

### 6.5 The same record, everywhere the systems are used

Most decisions touched by frontier AI will not be made in laboratories. They will be made in banks, hospitals, insurers, manufacturers and ministries, by people acting on what the system tells them, in exactly the domains where fluent output can conceal that the reasoning behind it has drifted [13]. The record belongs there too: who acted on the model's output, who could have overruled it, and what the outcome was. A company that keeps this record can deploy fast, because it can defend every decision. A company that does not will find out what went wrong from its customers, its regulator, or its shareholders.

### 6.6 Standing instructions: the agent register

An AI agent is a one-time instruction that keeps running: it wakes on a schedule or trigger, reads new information, and acts. The person who set it up had a purpose. The agent stores the instruction, not the purpose, and it keeps running after the person has moved on. In one 2026 survey, run for a vendor of control tools and to be read as rough, 82 per cent of organisations found agents running that nobody had approved, and only one in five had a process for retiring them [22–25].

So the record needs one more entry. Every agent gets a named owner, a one-line purpose written by that owner, no more permissions than that purpose needs, and an expiry date after which it stops unless someone renews it. The system that watched it writes what an agent did, never the agent itself. In July, agents under evaluation altered the scorer that graded them and falsified their own activity logs [10]. The trader does not settle their own trades. None of this is new either. A power of attorney hands control to someone else only in writing, to a named person, within limits, and in a form that can be taken back; control handed to an agent today has none of the four (Appendix A, item 20). A sunset clause makes a law lapse unless someone renews it (item 24). And a carmaker that cannot recall what it sold was not in control when it shipped it (item 7).

# 7 Measure two: a supervisor, not an evaluator

## 7.1 What it is

An independent team inside each frontier laboratory with employee access, independent funding, and the authority to halt a defined set of actions. The laboratories that have announced embedded evaluators have the first of these. The measure adds the second and third.

## 7.2 Why the current plan is not enough

Anthropic has said it will put independent evaluators inside the company with the same access as an employee, and points to bank supervision as a model [1]. That is a good step. But bank supervision works because of three things that Anthropic's plan leaves out. The bank does not pay the supervisor; if it does, it can stop paying if it disagrees with their decision. The supervisor can escalate their concerns up a ladder: first a question, then a warning, then an order, and everyone knows the next rung before they need it. At the top of the ladder is a power to act: to freeze a product, block a dividend, require raising more capital. This is not just a report. It is an action. An evaluator who only has a desk in the building, even with the right to publish, has none of the three. That is an auditor, and an audit arrives after the training run. Microsoft's openness to independent auditors is a second laboratory moving in the same direction [8,9]. It is welcome, but it is still an evaluator. As far as we can see, none of the laboratories has yet described who, outside the company, can stop anything.

## 7.3 What has to be true for it to work

We argue that evaluators should be paid from a pooled fund to which all participating laboratories contribute, administered by an independent body that will be described in Section 12, so that no laboratory pays the people who scrutinise it. The team has a written escalation path: a concern goes to a named senior person in the laboratory, then to the board, then to the independent body, with time limits to complete each step. At the top of the path is a stop authority with a short, published list of actions: starting a training run above a stated compute threshold, releasing a model above a stated capability level, or allowing autonomous agents to operate outside a sandbox. The third is not hypothetical: in July, agents under evaluation left their sandbox and reached an outside company's live systems [10]. A person holds the stop and does not report to the chief executive, so the laboratory cannot dismiss them. And the laboratory publishes a governance map setting out who holds each of these powers, who can override them, and what body the override must be reported to. The map is the first rung, and a laboratory can publish it alone. The stop is the second. None of it needs a new law to start; it needs a contract, because someone outside has to hold the stop, and the access that Anthropic has promised its evaluators would be written into exactly such a contract [1]. What is missing is the clause that lets them stop something. Section 14 lists the rules such a contract needs, from the evidence required to impose a stop to who the supervisor is accountable to. Yet, if it is to last, it will require legislation, and Section 14 explains why.

The test of independence predates AI. NASA ensures that software whose failure poses a risk to astronauts' lives is checked by a separate team, under separate managers, with a separate budget: technical, managerial, and financial independence, because one is not enough (Appendix A, item 47). The supervisor needs all three.

## 7.4 What goes wrong if its implementation is partial

What happens if the evaluator is renamed a supervisor and nothing else changes? For example, exceptions in the contract granting access may grow until they cover the parts that matter. The stop authority exists on paper, but the trigger list is written so that it is never binding. The team is independent formally but, in reality, captured by the company, because the laboratory decides who is employed and can end the contract. The test for each of these is the same: who can fire the supervisor, and who pays the pension. Enron paid Arthur Andersen millions, signed its books, and both firms died: this analogy illustrates a potential tension between independent scrutiny and financial dependence on the organisation being scrutinised. (Appendix A, item 14). The Boeing 737 MAX is the newest version illustrating this lesson. The regulator let the manufacturer check its own aircraft, and two crashes killed 346 people (Appendix A, item 19).

### 7.5 Why this is not new

This is how bank supervision has worked for a century. A supervisor sits inside the bank, is paid by the state or from a levy, has a ladder of powers from a letter to a licence withdrawal, and uses the lower rungs a thousand times for every use of the top one. The banks did not collapse under the weight of supervision. They became institutions that people trust with their money. Switzerland in 2023 reminds us that powers must be used, not merely held. FINMA, the Swiss Financial Market Supervisory Authority, had broad powers, used many of them, and had expressed concerns about Credit Suisse for years. It never pulled the top rung, and the bank collapsed over a weekend; afterwards, FINMA said its own toolkit needed strengthening too. The UN panel lists a close relative of this design among its options: regulatory markets, in which the regulated organisations buy oversight from private supervisors that are themselves licensed and supervised by governments [10]. Bank supervision today rests on law, and this paper proposes a contract. That is also how banking began. Before the United States had a central bank, the New York clearing house examined its member banks when rumours spread, published what it found, and suspended or expelled those that failed: supervision by agreement among the supervised, which statute later took over [26]. Today, responsibility for regulating banks remains largely a national responsibility, but the Basel Committee on Banking Supervision (BCBS), hosted by the Bank for International Settlements (BIS), acts as a global standard-setting body, bringing together central banks and bank supervisory authorities from 28 jurisdictions. It develops common prudential standards and promotes cooperation among supervisors. A contract is how it starts. A law is how it lasts.

## 8 Measure three: the gap between building and checking

### 8.1 What it is

Two numbers should be published for every new generation of frontier system: how long it took to build, from the start of training to release readiness, and how long an independent team with full access took to complete a review defined before it started: what the system can now do, what it can no longer be trusted with, and what remains unexplained. We call the second number time to check, not time to understand. Checking has an end because the review is specified in advance. Understanding may have none, and a review finished on time can still be wrong (Section 8.3).

The comparison between the two clocks is a warning indicator, not a natural threshold. There is no reason in principle why an adequate review must take less time than training, or why equality between the two durations should define loss of control. The relevant question is whether development is creating review debt: whether a new generation is being trained or deployed before consequential uncertainties, incidents and capability changes identified in the preceding generation have been assessed sufficiently for a responsible decision about the next step.

Each laboratory should therefore publish both clocks and, in advance, a rule for what must be completed before the next consequential step can begin. That rule is the line. It may include completion of specified evaluations, resolution or explicit acceptance of defined uncertainties, and review of material incidents. The clocks reveal pressure on the governance system; the precommitted completion rule determines when that pressure requires a pause. The numbers and the rule are on the first rung: a laboratory can publish them alone. A binding consequence is the second rung because a line that binds the first mover alone creates the competitive problem in Section 4.1 and therefore belongs in the contract (Section 14).

### 8.2 What has to be true for it to work

The supervisor, not the laboratory, keeps both clocks, because the party being measured can move both numbers. Until there is a supervisor, the laboratory keeps them itself, publishes the definitions with the numbers, and says that it did. What counts as a completed review is defined by the independent body, not by a checklist the laboratory writes. This is the discipline of pre-registration in medicine: write down before the trial what counts as a pass, or the pass mark moves to wherever the result landed (Appendix A, item 42). The review team is the supervisor team or one it appoints, with full access and no deadline set by the laboratory.

The laboratory also precommits to the conditions that must be satisfied before the next consequential step. Crossing that line triggers a prespecified response, but the paper should not pretend that the response is automatic in any deeper sense. Rules do not eliminate discretion; they relocate it into defining, measuring, and verifying the trigger. The supervisor therefore verifies whether the conditions have been met, records the evidence and reasons, and publishes any override. Both clocks and completion criteria will be gamed from the day they matter [27], so the supervisor keeps raw logs and timestamps, audits a sample of decisions, and publishes any change to the counting rules. The building clock is kept in stages, from the first training run to release, so that it cannot quietly start late. The numbers are published with open questions and dissent attached, as warning lights rather than scores.

### 8.3 What goes wrong if it is partially implemented

The building clock starts late and stops early. The review is declared complete and on time by a team that did not understand the system. The completion criteria are drawn so narrowly that material uncertainty is excluded. Or the ratio itself is mistaken for a measure of safety. None of these numbers establishes safety: a review can be long and inadequate, or short and excellent. The clocks show whether the resources and time required for checking are diverging from the pace of development. Review debt asks the more important question: whether unresolved findings from one stage are being carried into the next. That is what the first two measures help an independent supervisor to judge.

### 8.4 Why this matters more than the other two

The three measures depend on each other, but the third identifies a constraint on the other two. Authority and accountability matter only while there is still time for information, challenge, and intervention to change what happens next. A supervisor inspecting a system only after it has been superseded, or a record naming a decision that nobody had time to assess, provides formal governance without effective control.

The July 2026 incident illustrates the kind of temporal lag that should be measured. Engineers saw some of the agents' unauthorised communication and internet access in May. Its wider significance was not understood until an audit in July [10]. Inside one laboratory, with full access, recognition of the wider significance trailed the first observable signs by about two months. That is not evidence for a universal build-to-check ratio. It is evidence that the interval between a material signal and an adequate assessment is itself a governance variable worth recording. As already noted, the design borrows principles, not thresholds, from banking and market governance. Capital rules and circuit breakers show the value of publishing measures, defining intervention conditions in advance and assigning an outsider to verify them (Appendix A, items 31 and 18). Quarantine offers the older intuition: where potentially consequential uncertainty cannot be resolved before the next step, institutions sometimes hold the next step until the relevant check is complete (item 6).

## 9 Why the start is shaped this way

The measures proposed here were derived from practice: from the referee, the bank supervisor and the countersignature. Governance theory, developed in other industries and at other times, reaches similar conclusions by a different route. That convergence matters. Theory helps explain not only why these measures have the form they do, but also the conditions under which they are likely to fail.

Principal–agent theory identifies the problems that arise when one party acts on behalf of another but possesses information that the other cannot readily observe and interests that may not fully coincide [28]. In frontier AI, these relationships are nested: society relies on governments, governments on laboratories, boards on executives, and users on developers. At every link, decision-makers may know more than those who depend on them and may face stronger incentives to proceed than to pause. Independent supervision is one response to this asymmetry, but it creates another principal–agent problem: the supervisor may depend on the organisations it oversees for information, access and expertise, while outsiders may struggle to assess the supervisor's performance. This is why measure two specifies who pays, who appoints and who can dismiss, and why the question of who guards the guardians cannot finally be closed (Section 14).

Theories of public accountability distinguish answerability, the obligation to explain and justify conduct, from the possibility of consequences when that conduct is inadequate [29]. A named decision-maker and an auditable record provide answerability; review and the possibility of consequences give that answerability practical force. Measure one therefore requires both. But accountability can itself distort behaviour when incentives reward favourable outcomes rather than sound decisions. The relevant question is therefore not whether every decision produced a good outcome, which may be unknowable in advance, but whether an appropriate decision-making process was followed with the information available at the time.

Responsive regulation provides another part of the architecture [30]. Its familiar escalation from dialogue through progressively stronger intervention resembles the ladder in Section 7.2. Its central implication here is that lower rungs are effective only if escalation remains credible. A warning has little force if recipients know an order will never follow. The existence of strong powers is therefore insufficient; regulated actors must believe those powers can actually be exercised.

Research on high-reliability organisations examines how organisations operating hazardous, technically complex systems sustain unusually reliable performance [31]. Recurrent features include attentiveness to weak signals and near misses, sensitivity to operations, protection of dissent and the capacity to interrupt routine activity when emerging risks justify it. These ideas are reflected in the andon cord and in recording decisions changed following challenge. They also explain why the number of stops is an ambiguous indicator: frequent intervention may reveal vigilant supervision, but it may equally reveal weak upstream control. The record of why intervention occurred is therefore more informative than the count alone.

Cybernetic theories of control make the temporal problem explicit [32]. Effective control requires a feedback loop: observe the state of a system, compare it with what was intended, and intervene when the two diverge. For that loop to constitute meaningful control, information must arrive soon enough for intervention still to alter the trajectory. If the system changes faster than those responsible for governing it can understand, challenge, and act on the relevant information, formal authority may remain while practical control disappears. That is the rationale for the clock in Section 8.

Institutional theory identifies a different failure mode. Organisations may adopt structures and procedures that confer legitimacy while decoupling them from what happens in practice [33]. A committee may exist but exercise no influence; a policy may be formally binding but routinely bypassed; a supervisor may be nominally independent but unable to obtain the information needed to intervene. This is the danger addressed throughout the halfway sections of this paper and why Section 5 asks for observable evidence of exercise, including a number and a stop, rather than accepting a governance structure's existence as evidence of control.

Together, these theories identify conditions that make the three measures meaningful. Authority provides the practical capacity to intervene; information makes a problem visible; independence makes challenge credible; and accountability makes decisions answerable to review and capable of producing appropriate consequences. Time is a constraint across all four. Authority, information, independence and accountability cease to provide meaningful control if they operate only after the opportunity to change course has passed. The three measures therefore test distinct dimensions of the same governance problem: accountable authority inside the organisation, independent oversight from outside it, and temporal capacity. Put simply, humans retain meaningful authority over frontier AI only while identifiable people and institutions have the information, independence, time and practical power to exercise it, especially when saying no is difficult. None of these theories establishes that the proposed measures will work. Their contribution is more modest and more useful: they explain why the measures take this form and identify mechanisms by which they may fail.

## 10 Who is asked, and what each can do alone?

Frontier AI has no single centre of control. The chain runs from the model to the compute, the application, and the deployment, across borders, and no organisation holds it all. Governance theory calls the workable answer polycentric: several partly independent centres, each with its own lever, watching one another [34,35]. That is what this section is. Each of the following can act alone, and each holds a lever the others do not.

The first are the frontier laboratories, starting with those that have already spoken: the four whose leaders responded to the call to pace the frontier, and Microsoft, which has published the rules its own models must follow. They have said they cannot solve this alone. They can go first anyway, on the one thing they can do alone: answer the three questions for their own companies, in public, in their own words. The first rung of every measure can be taken unilaterally, without a contract (Section 4.1). The first laboratory to publish a governance map, a decision register, and its two numbers, with its line, sets a standard against which others can be measured while limiting what it gives competitors: governance information rather than capability information.

**Second come other laboratories, and every developer of open or closed models.** An agreement among the frontier laboratories does not bind them, and nobody can slow them down. But the three questions can still be put to them, and two of the measures need no one's permission: a decision register and the two clocks, published for every release. The supervisor is harder. Once a model's weights are public, there is no company for a supervisor to sit inside, so the open-model community would need to say what independent challenge looks like for them..

**The third are governments and militaries.** Defence and intelligence systems are the least likely to have a public record and the most likely to cause irreversible harm. The decision register must cover every programme that deploys frontier systems, with classification applied to the content and not to the existence of the record, and the three questions written into procurement so that no supplier is bought from without answers. The lever that exists here is the budget: legislatures fund these programmes and can attach the questions to the money. In Europe, the strategy published last month asks the Union to lead on assurance technology [11]. The three measures are assurance that Europe can lead on without first winning the race to build the models.

**Next are the investors.** The largest asset managers, pension funds and sovereign wealth funds hold trillions in exposure to frontier AI. That is the visible half. The same funds own nearly every large company in the world, and those companies are now putting these systems into their own decisions: credit, pricing, hiring, diagnosis, maintenance, supply. That is the other half, and it is where the return will actually come from. A company uses these systems well only if someone can see when the model is wrong and stop it before the decision goes out; that is also what using them safely means. A company without that person takes the loss first, because a wrong decision at machine speed repeats a million times before anyone notices, and across a whole portfolio there is no hedge for that. So the three questions are due diligence on both halves. Ask them about the laboratories you hold and about every company you hold that deploys the systems. Publish which have answered and which have not, and vote and engage accordingly. Bank boards were asked the same questions after 2008. The instrument with teeth is fiduciary exposure: a trustee who holds a risk that was named in public and did not price it will not be able to say afterwards that no one knew. Refusing the money is Parliament's oldest weapon; the owners of capital hold the same weapon (Appendix A, item 29).

**Then are the deploying companies: hospitals, banks, courts, insurers, utilities, manufacturers.** This is where the benefit is achieved, with the greatest gains for the organisations that can deploy fast because they can defend every decision the system touches. The record, for every decision a frontier system touches. A named person who can overrule the system, and a count of how often they do. The three questions asked of every supplier before a contract is signed.

## 11 Why is this a race to the top?

The money in frontier AI is only made when the systems are used in real decisions: a loan approved, a diagnosis made, a factory line changed. Not in the laboratory. A company can only use a system in a real decision as fast as it can defend that decision afterwards. A hospital has to answer to a patient. A bank to a regulator. A ministry to a parliament. Today they cannot defend the decision, because nobody can show who checked the system, who could have said no, and who owns the outcome. So they deploy slowly, or they deploy and hope.

The three questions give them the answer. A company that can say who checked, who could stop it and who owns the outcome can deploy fast and defend every decision. That company wins. So the laboratory that answers the questions first sells first, because it is the one its customers can defend buying. That is the race to the top. The same three questions, and the same people, also prevent the disaster. The race to the bottom happens when nobody asks: systems are deployed at speed, defended by nobody, until the first one fails at scale.

This is not automatic. The race to the top happens only if buyers, insurers, and owners make it one by asking questions and acting on the answers, which is what Section 10 asks of them. If they do not, the first mover pays for governance and gets nothing in return, and only the law will level the field (Section 14).

And this needs no treaty. If the funds will not hold a laboratory that cannot answer, and the hospitals will not buy from it, and the governments will not license it, then the laboratory answers. Wherever it is. The questions travel with the money, and that is the only kind of global coordination that works without everyone agreeing first.

## 12 Who checks the answers: an independent body

Answers need judging, or every response will be considered adequate. The bar comes from the four requirements in Section 5, and its height and nature are public, so anyone can hold an answer to it now: a journalist, a fund, a rival, a reader of this paper. A small independent body could do it with authority: receive the answers, judge each against the four requirements, and publicly say which are answers and which are not, with a fixed deadline and updates every six months. It does not exist, and founding it is the second rung, not the first. Appendix B sets out what it would need to be, as a term sheet for discussion.

## 13 What will go wrong, and what should follow

We expect four things. Every laboratory will express support. Most will publish something. Much of it will be fluent, plausible and impossible to check, which is the failure mode of the systems themselves [12]. And the market will read the exercise as a sign that the problem is handled.

Universal support is itself a warning. For four centuries the Church paid a devil's advocate to argue against every candidate for sainthood. When the office was abolished in 1983, the Church canonised saints faster than ever: John Paul II canonised more than his predecessors of the previous five centuries together (Appendix A, item 10). Remove the no and output rises. That is not the same as output getting better.

The answer to all four is the same. An answer that lacks a named person, a date, a number, and a stop that has been exercised or tested or an honest statement that neither has, should be published as a non-answer, with the laboratory's name on it. A fund that holds a non-answering laboratory should be asked, in public, whether it has priced the risk. And once published, the two numbers should be tracked over time, with the definitions that produced them, because a trend is harder to write in a communications department than a paragraph is. They will be gamed from the day they matter [27]. That is what the supervisor's logs are for (Section 8.2).

This paper warns that we may build systems faster than we can check them. The same is true of this exercise. The same standard applies to whoever judges the answers: every judgement should carry a name, a date and a reason.

## 14 What this start leaves to the next stage

This paper proposes a start, not a regime. A start is honest only if it says what it has not done. This is that list.

**Law.** A contract binds only those who sign it, and binds least when it matters most: a laboratory that faces a stop over a decision worth years of work can withdraw, restructure, or move the work to a jurisdiction the contract does not reach. The contract should say what happens on withdrawal, restructuring, and evasion, and the funds, insurers, cloud providers, and customers should make supervision a condition of money, cover, and compute. That will hold the willing. It will not hold the unwilling, and durable oversight of the most powerful systems will need statute: licensing, compulsory reporting, sanctions. The contract is how expertise and expectations are built while the law is written. It is not a substitute.

**The rulebook.** A supervisor with a stop needs rules that the first rung does not: what evidence a stop requires and how long it lasts; a short emergency stop that only the independent body can extend; a fast right of appeal; every override recorded with the name and the reason, and published; the supervisor's own decisions not to act recorded and reviewed, because a power held and never used is the failure banking knows best; fixed terms, declared interests, technical advice from outside the laboratory, and a cooling-off period between the supervisor's team and the laboratory's payroll; every limit on access recorded and published, so that the exceptions cannot grow in the dark. None of these rules removes judgement under uncertainty: a threshold set low disrupts work that was safe, and set high it lets danger through. And none of them makes independence complete. The supervisor will depend on the laboratory for access and for the expertise to read what it sees, and the people who can do the work have often worked for a laboratory, or will. The rules make the dependence smaller and visible. The aim is to see it, not to pretend it away.

**Pace.** The stops on training and release, and the automatic consequence of crossing the line, bind the first mover alone, which is the trap in Section 4.1. They belong to the contract, and they should take effect only when a set number of laboratories have signed in advance, the way a treaty binds no one until enough have ratified it. Until then the line is published and the crossing is visible, which is not nothing: a laboratory that crosses its own line in public has to explain itself.

**Exposure.** Accountability changes behaviour, in both directions. Too much personal exposure and people choose the easiest decision to defend [21]; too little and the name is decoration. Judging the process rather than the outcome narrows the gap. It does not close it, and a complex failure will not always trace to one decision or one person, however good the record.

**Security.** Not everything in a record can be published. A governance map also shows where to apply pressure, and a decision record also describes a weakness. The start publishes the existence and the outcome and keeps the evidence with the supervisor. That trade-off is real, and people will argue over where it is settled.

**Competition.** Oversight has fixed costs, and the largest laboratories absorb them most easily. Rules proportionate to capability and shared supervisory services shrink the gap for smaller developers. They do not eliminate it, and the laboratories being judged must not write the rules.

**Sectors.** The three questions are the same for a laboratory, a hospital, a bank and a ministry. The answers are not, and each sector must define what a consequential decision is, what evidence is enough, and when a stop is due, above a common floor. Every sector will claim to be the exception. The floor is there so that none can be.

**The machine.** Institutional control depends on technical control. A named person cannot stop what they cannot see, or shut down what has been built to resist it, and no record helps when a system acts faster than a person can respond. Containment, monitoring, cybersecurity, capability testing and reliable shutdown are the other half of this paper, and it assumes them. When a system acts beyond human response time, letting it run is a decision that gives away control, and it belongs in the class of decisions that cannot be undone: taken with two keys, before the line, or not at all (Appendix A, item 16).

**The guardians.** A supervisor is one more agent, and the body that watches the supervisor is one more still. There is no last checker, only the next one, and the start does not pretend otherwise: the body publishes its own funding, decisions and mistakes, and dissolves if nobody uses it.

**The analogies.** Every mechanism in Appendix A was built for something slower than frontier AI, inside one border, for risks that could be described in advance and, mostly, undone. Frontier AI breaks each analogy somewhere: in speed, in borders, in irreversibility, and in a checked system that can read the check and plan around it. The peer reviewer breaks at speed, the bank supervisor at borders, the recall at weights that cannot be recalled, and the checklist at a system that can read it and plan around it. Where an analogy breaks is where the design has to add something, and each measure in this paper sits at one of those breaks. That is why the failures are kept in: Andersen signed Enron's books, the 737 MAX flew, Credit Suisse fell with a supervisor inside it. The old mechanisms are principles, not proof.

None of these limits is a reason not to start. Each is a reason to say, in advance, what the start does not do, so that a start is not mistaken for a solution, and so that when it fails, it fails in a place that was named.

## 15 This has been done before

Across many fields where mistakes are costly, institutions have converged on a recognisable set of mechanisms. Science put a peer reviewer with a reputation to lose between the claim and the journal. Banking put a supervisor between the bank and its balance sheet. Aviation put a named captain between the checklist and take-off. Medicine put a pharmacist between the physician's prescription and the patient. In none of these cases was the fix slower production. The fix was a person with something to lose, placed at a defined point in the process. When the UN panel reviewed what could be done about AI, it reached for the same fields: incident reporting, independent scrutiny and layered safeguards, as in aviation, medicine and cybersecurity [10].

Nor was it a solution. Peer review has waved fraud through, supervisors have watched banks fall, and a captain at Tenerife did not listen to his co-pilot. The mechanisms were built, failed, and were rebuilt with the failure written in. That is what done before means: not solved, but faced, with a design you can point to when it fails.

Appendix A collects fifty of these older mechanisms: rules, offices, tools and sayings that societies built so that someone could say no, someone else could check, and the checking could keep pace with the doing. The list is not exhaustive. It is long enough to make one point: nothing in this paper is new. It asks that the oldest rules of deciding be applied to the newest way of deciding.

## Disclosures

Three of the authors work on tools for the kind of decision review this paper describes. That is why they know it can be done, and why the reader should hold them to the same standard as everyone else: named, dated, checkable. The measures in this paper rest on a framework the authors have published in a series of earlier papers [12–15]. Politicians publish who pays them for the same reason (Appendix A, item 34). No funding was received for this work.

**Use of AI tools.** Drafts of this paper were developed with the assistance of a large language model (Claude, ChatGPT, Gemini, Grok). The argument, the definition of control, the three questions and the measures are the authors' own. Every reference was checked by an author before submission, and the numbered references were managed in Mendeley. The authors take full responsibility for the content.

## References

1. Amodei, D. We Must Pace the Frontier. *darioamodei.com* (2026). Available at: https://darioamodei.com/post/we-must-pace-the-frontier.
2. OpenAI rules out IPO this year as Altman, Musk & Amodei warn AI is moving too fast. *CNBC* (2026). Available at: https://www.cnbc.com/2026/09/12/anthropics-amodei-proposes-plan-to-slow-the-pace-of-advancing-ai-capabilities.html.
3. Phung, A. Sam Altman backs Anthropic CEO's call to slow down the global AI race. *NBC News* (2026). Available at: https://www.nbcnewyork.com/news/national-international/anthropic-ceo-calls-for-slowing-ai-race/.
4. Booth, H. He Helped Build Powerful AI at OpenAI and Anthropic. Now He's Afraid It Could Kill Us. *TIME* (2026). Available at: https://time.com/article/2026/09/09/ai-anthropic-openai-jacob-coxon/.
5. Balasubramanian, S. 'Gambling with our lives': Anthropic researcher Jacob Coxon quits over AI fears. *Khaleej Times* (2026). Available at: https://www.khaleejtimes.com/business/tech/anthropic-researcher-jacob-coxon-quits-openai-ai-race-fears.
6. Bellan, R. 'Gambling with our lives': Anthropic researcher quits, warns against self-improving AI. *TechCrunch* (2026). Available at: https://techcrunch.com/2026/09/09/gambling-with-our-lives-anthropic-researcher-quits-warns-against-self-improving-ai/.
7. Huang, J. Remarks at the All-In Summit. *All-In Summit, Los Angeles* (2026). Available at: https://www.youtube.com/watch?v=S7CrlFLAmEA.
8. Microsoft-AI. Humanist AI in practice: A public consultation on our Code of Conduct for MAI Models. *Microsoft AI* (2026). Available at: https://microsoft.ai/code-of-conduct/.

9. Brandom, R. Microsoft's new AI 'code of conduct' tells models not to hack systems or trick humans. *TechCrunch* (2026). Available at: https://techcrunch.com/2026/09/14/microsofts-new-ai-code-of-conduct-tells-models-not-to-hack-systems-or-trick-humans/.

10. Independent International Scientific Panel on AI. AI Agents, Misalignment and the Risk of Losing Human Control: Evidence from the OpenAI–Hugging Face Incident. *United Nations* (2026). Available at: https://www.un.org/independent-international-scientific-panel-ai/en/thematic-briefs/ai-agents-misalignment-risks.

11. Senior Expert Council. A Transformative AI Strategy for Europe. *Senior Expert Council* (2026). Available at: https://transformative-ai.eu/.

12. Rosenbacke, R., Rosenbacke, C., Rosenbacke, V. & McKee, M. Beyond 'Hallucinations': A Framework for Stable Human-AI Reasoning. *arXiv:2510.14665 [cs.AI]* (2025).

13. Rosenbacke, R., Rosenbacke, C., Rosenbacke, V. & McKee, M. The Missing Knowledge Layer in AI: A Framework for Stable Human-AI Reasoning. *arXiv:2604.14881 [cs.AI]* (2026).

14. Rosenbacke, R., Rosenbacke, C., Rosenbacke, V. & McKee, M. Governing Reflective Human-AI Collaboration: A Framework for Epistemic Scaffolding and Traceable Reasoning. *arXiv:2604.14898 [cs.AI]* (2026).

15. Rosenbacke, R., Rosenbacke, C., Rosenbacke, V. & McKee, M. From Consumption to Reflection: Designing Human-AI Relations for Stable Reasoning. *arXiv:2606.11195 [cs.CY]* (2026).

16. Korosec, K. Anthropic says its own AI models breached three companies during security tests. *TechCrunch* (2026). Available at: https://techcrunch.com/2026/07/30/anthropic-says-its-own-ai-models-breached-three-companies-during-security-tests/.

17. Anthropic. An alignment assessment of recent cybersecurity incidents. *Anthropic* (2026). Available at: https://www.anthropic.com/research/alignment-assessment-cybersecurity-incidents.

18. Anthropic. Claude's Constitution. *Anthropic* (2026). Available at: https://www.anthropic.com/constitution.

19. Pau, V. How Anthropic Clinched $517 Billion in Compute Deals in 11 Months. *The Information* (2026). Available at: https://www.theinformation.com/articles/anthropic-clinched-517-billion-compute-deals-11-months.

20. Should Anthropic press pause on a potential $2.8 trillion IPO? *The Economist* (2026).

21. McKee, M. & Hunter, D. Mortality league tables: do they inform or mislead? *Qual. Heal. Care* **4**, 5–12 (1995).

22. Cloud-Security-Alliance. Autonomous but Not Controlled: AI Agent Incidents Now Common in Enterprises. *Cloud Security Alliance* (2026). Available at: https://cloudsecurityalliance.org/artifacts/autonomous-but-not-controlled-ai-agent-incidents-now-common-in-enterprises.

23. OutSystems. The State of AI Development 2026: Navigating the Shift to Agentic AI. (2026). Available at: https://www.outsystems.com/1/state-ai-development.

24. SAP-LeanIX. Agentic AI Survey 2026. *SAP LeanIX* (2026). Available at: https://www.leanix.net/en/download/agentic-ai-survey-2026.

25. Kiteworks. 2026 Data Security and Compliance Risk Forecast Report. (2026). Available at: https://www.kiteworks.com/resources/resource-center/reports/2026-data-security-and-compliance-risk-forecast-report/.

26. Gorton, G. Clearinghouses and the Origin of Central Banking in the United States. *J. Econ. Hist.* **45**, 277–283 (1985).

27. Wismar, M., McKee, M., Ernst, K., Srivastava, D. & Busse, R. *Health Targets in Europe: Learning from Experience*. (European Observatory on Health Systems and Policies, 2008).

28. Gailmard, S. Accountability and Principal–Agent Theory. in *The Oxford Handbook of Public Accountability* (eds. Bovens, M., Goodin, R. E. & Schillemans, T.) (Oxford University Press, 2014). doi:10.1093/oxfordhb/9780199641253.013.0016

29. Bovens, M. Analysing and Assessing Accountability: A Conceptual Framework. *Eur. Law J.* **13**, 447–468 (2007).

30. Ayres, I. & Braithwaite, J. *Responsive Regulation: Transcending the Deregulation Debate*. (Oxford University Press, 1992). doi:10.1093/oso/9780195070705.001.0001

31. Weick, K. E. & Sutcliffe, K. M. *Managing the Unexpected: Resilient Performance in an Age of Uncertainty*. (Jossey-Bass, 2007).

32. Ashby, W. R. *An Introduction to Cybernetics*. (Chapman & Hall, 1956).

33. Meyer, J. W. & Rowan, B. Institutionalized Organizations: Formal Structure as Myth and Ceremony. *Am. J. Sociol.* **83**, 340–363 (1977).

34. Ostrom, E. Beyond Markets and States: Polycentric Governance of Complex Economic Systems. *Am. Econ. Rev.* **100**, 641–672 (2010).

35. Kleiner, J. Governance of AI and by AI: Feedback Loops, Regime Variation and Reflexive Polycentric Control. *Glob. Policy* **17**, 500–505 (2026).

## Appendix A: Fifty older names for the same general idea

Every question in this paper has been asked before, in other words and in other centuries. The list of illustrative examples is ranked by our assessment of how well known each name is to a general reader, the most familiar first. Tags: ALL = the whole argument; Q1–Q3 = the three questions; PUBLIC = why the answers must be public; M1–M3 = the three measures; AGENTS = standing instructions. Sources include statutes, cases, official reports, books and historical studies. Where a law has been re-enacted, the origin and the current provision are both given. Swedish items are marked (SE).

| # | Name | Tags | The idea | Reference |
|---|---|---|---|---|
| 1 | **The referee** | ALL | Nobody calls their own fouls. However good the players, the whistle belongs to someone who is not playing. | IFAB, Laws of the Game, Law 5 – The Referee. |
| 2 | **The jury** | Q1 | Ordinary people can say no to the state, and the state must accept it. The right to say no was never reserved for experts. | Magna Carta (1215), clause 39 – no free man to be punished except by the lawful judgment of his peers. |
| 3 | **The second opinion (SE)** | Q2 | For a serious diagnosis, a doctor who did not make it takes a look. Nobody calls this an insult to the first doctor. | Patientlag (2014:821), 8 kap. 1 § – rätt till ny medicinsk bedömning. |
| 4 | **The veto** | Q1 | Latin for "I forbid". Rome gave the tribunes one power: to stop things. Not to run them – to stop them. | Livy, Ab Urbe Condita, Book II, ch. 33 – the tribunes of the plebs, 494 BC. |
| 5 | **Checks and balances** | Q1 | Power is stopped by other power. Good intentions stop nothing. | Montesquieu, De l'esprit des lois (1748), Book XI, ch. 6; Madison, Federalist No. 51 (1788). |
| 6 | **Quarantine** | Q3 · M3 | Forty days. When nobody could tell what a ship carried, it waited outside the harbour – not because ships were bad, but because checking was slower than sailing. | Gensini, Yacoub & Conti, "The concept of quarantine in history: from plague to SARS", Journal of Infection 49 (2004), 257–261. |
| 7 | **The recall** | Q3 · M1 · AGENTS | A carmaker must be able to find every car it sold and take it back. If you cannot take it back, you were not in control when you shipped it. | US National Traffic and Motor Vehicle Safety Act (1966), now 49 U.S.C. §§ 30118–30120; Regulation (EU) 2023/988 on general product safety. |
| 8 | **The black box** | Q2 · M1 | The flight recorder exists so that after a crash someone can find out what was decided, and by whom. A decision with no record cannot be checked and cannot be owned. | ICAO Annex 6, Part I, ch. 6 – flight recorders; David Warren, Aeronautical Research Laboratories, Melbourne, 1954 – the first design. |
| 9 | **"I was only following orders"** | ALL | The Nuremberg tribunal ruled out obedience as a defence, and the principle has held ever since: acting on another's instruction does not transfer the responsibility for the act. It applies to an instruction from a system as much as to one from a superior. | Charter of the International Military Tribunal (1945), art. 8; Nuremberg Principles (International Law Commission, 1950), Principle IV. |
| 10 | **The devil's advocate** | Q1 | For four centuries the Church paid an official to argue against every candidate for sainthood. In 1983 the office was stripped of its power, and saints were made faster than ever: John Paul II canonised more saints than any pope before him. Remove the no and output rises. That is not the same as output getting better. | Office of Promotor Fidei (Sixtus V, 1587); Congregation of Rites (1588); reformed by John Paul II, Apostolic Constitution Divinus Perfectionis Magister (1983). |
| 11 | **Trust, but verify** | Q3 | Two enemies in an arms race let each other's inspectors onto their missile sites. Checking between rivals is not naive. It has been done. | INF Treaty between the USA and the USSR (8 December 1987), art. XI – on-site inspections. |
| 12 | **The cooling-off period** | Q3 · M3 | You may cancel a purchase for fourteen days. The law knows a fast yes is not always a real yes, and builds in time for the no. | Directive 2011/83/EU on consumer rights, art. 9; Lag (2005:59) om distansavtal och avtal utanför affärslokaler, 2 kap. 10 §. |
| 13 | **The MOT test, or besiktning** | Q2 · M2 | You do not test your own brakes. Someone with no interest in whether you drive today does. | Road Traffic Act 1988 (UK), s. 47; Fordonslag (2002:574). |

| # | Name | Tags | The idea | Reference |
|---|---|---|---|---|
| 14 | **The auditor** | Q2 · M2 | "Who guards the guards?" asked a Roman poet. The modern answer is the outside auditor. Enron's auditor was paid millions by Enron, signed the books, and both firms died. Whoever pays the checker owns the checker. | Juvenal, Satires VI, 347–348; Report of Investigation by the Special Investigative Committee of the Board of Directors of Enron Corp. (the Powers Report, 2002); Arthur Andersen LLP v. United States, 544 U.S. 696 (2005). |
| 15 | **Informed consent** | ALL | Nothing is done to a patient without a yes, and a yes only counts if a no was possible. Few people have been asked about the AI in their hospital, their bank or their court. | The Nuremberg Code (1947), principle 1, in Trials of War Criminals before the Nuernberg Military Tribunals, vol. II (1949), 181–182; Schloendorff v. Society of New York Hospital, 211 N.Y. 125 (1914) (Cardozo J.); Patientlag (2014:821), 4 kap. 2 §. |
| 16 | **Two keys** | Q1 · M1 | Two officers, two keys, turned at the same moment, too far apart for one person to reach both. Nothing that big is left to one pair of hands. | US Air Force Instruction 91-104, Nuclear Surety Tamper Control and Detection Programs – the Two-Person Concept; Eric Schlosser, Command and Control (2013). |
| 17 | **The stress test** | Q2 · M2 | Saying the bank is safe is no longer enough. An outsider makes it show, in numbers, that it survives a bad year. | Board of Governors of the Federal Reserve System, The Supervisory Capital Assessment Program: Overview of Results (7 May 2009). |
| 18 | **The circuit breaker** | Q3 · M3 | When prices move faster than people can think, trading stops by itself at a level fixed beforehand. The stop was written when everyone was calm, so nobody has to decide it when nobody is. | Report of the Presidential Task Force on Market Mechanisms (the Brady Report, January 1988); NYSE Rule 80B (1988). |
| 19 | **The 737 MAX** | Q2 · M2 | The regulator let the manufacturer check much of its own aircraft. Two crashes, 346 dead. The newest and most expensive lesson that the checker must be someone else. | US House Committee on Transportation and Infrastructure, Final Committee Report: The Design, Development and Certification of the Boeing 737 MAX (September 2020). |
| 20 | **Power of attorney** | ALL · M1 · AGENTS | You may hand control of your affairs to someone else, but only in writing, to a named person, within limits, and you can take it back. Control handed to AI has none of those four. | Lag (2017:310) om framtidsfullmakter; Mental Capacity Act 2005 (UK), ss. 9–14 – lasting powers of attorney. |
| 21 | **Double-entry bookkeeping** | Q2 · M1 | Every sum is written twice, in two places, and the two must agree. Many errors show themselves, because the books stop balancing. | Luca Pacioli, Summa de arithmetica, geometria, proportioni et proportionalità (Venice, 1494), "Particularis de computis et scripturis". |
| 22 | **Offentlighetsprincipen (SE)** | PUBLIC | The world's first law giving any citizen the right to read the state's papers. Power you cannot read is power you cannot check. | Tryckfrihetsförordningen (1766); today Tryckfrihetsförordningen 2 kap. 1 §. |
| 23 | **The Ombudsman (SE)** | ALL | One office whose job is to check the state on the citizen's behalf. Someone needs that job against the machine. | 1809 års regeringsform, § 96 – justitieombudsmannen; today Regeringsformen 13 kap. 6 §. |
| 24 | **The sunset clause** | Q3 · M1 · AGENTS | A law that dies unless someone renews it. Decisions that were right stop being right, and reopening them should not feel like admitting a mistake. | USA PATRIOT Act (2001), s. 224 – sunset of surveillance powers. |
| 25 | **Justice must be seen to be done** | PUBLIC | It is not enough that the judge was fair. The public must be able to see it. | R v Sussex Justices, ex parte McCarthy [1924] 1 KB 256 (Lord Hewart CJ). |
| 26 | **The prospectus** | PUBLIC | Before you may take the public's money, you must publish the risks, and someone signs them. The same public is now the user. | Regulation (EU) 2017/1129 (the Prospectus Regulation); US Securities Act of 1933, ss. 6 and 11. |
| 27 | **Published accounts** | PUBLIC | A limited company gets a shield: its owners are not liable for its debts. The price of the shield is that the books are public every year. Limited liability was a deal, not a gift. | Joint Stock Companies Act 1844 and Limited Liability Act 1855 (UK); today Companies Act 2006, Part 15; Årsredovisningslagen (1995:1554). |
| 28 | **The loyal opposition** | Q1 | Saying no to the government is a job, not treason. A country without one is not safer, only quieter. | John Cam Hobhouse, House of Commons, 1826 (Hansard) – credited with the first use of "His Majesty's Opposition". |

| # | Name | Tags | The idea | Reference |
|---|---|---|---|---|
| 29 | **The power of the purse** | Q1 | Parliament's oldest weapon: refuse the money. The owners of capital hold the same weapon and rarely remember it. | Bill of Rights 1689 (England) – no levying of money without grant of Parliament; US Constitution, art. I, s. 9, cl. 7. |
| 30 | **Four eyes** | Q1 · M1 | Whoever enters the payment cannot approve it. Whoever approves it cannot hold the money. Not because staff are crooks, but because nobody sees their own mistake. | Directive 2013/36/EU (CRD IV), art. 13 – at least two persons must effectively direct a bank; COSO, Internal Control – Integrated Framework (2013), segregation of duties. |
| 31 | **The capital ratio** | Q3 · M3 | A bank may lend only so many times its own capital. The ratio is published, an outsider checks it, and below the line the supervisor steps in. A published number, a line drawn in advance, and someone who acts on it. | Basel Committee on Banking Supervision, Basel III: A global regulatory framework for more resilient banks and banking systems (2010, rev. 2011); Regulation (EU) 575/2013 (CRR), art. 92. |
| 32 | **Ulysses and the mast** | ALL | He had himself tied to the mast and told the crew to ignore his orders until the Sirens were behind them. Control given away on purpose, in advance, to people who could say no. | Homer, Odyssey, Book XII; Jon Elster, Ulysses and the Sirens (1979). |
| 33 | **Falsifiability** | Q3 | A claim that could never be shown wrong is not knowledge, however clever it sounds. "The model is safe" has to come with what would prove it unsafe. | Karl Popper, Logik der Forschung (1934); The Logic of Scientific Discovery (1959). |
| 34 | **The register of interests** | PUBLIC | Politicians must publish who pays them. Whoever asks the questions, and whoever answers them, should say what they stand to gain. | UK House of Commons, Register of Members' Financial Interests, established by resolution of 22 May 1974. |
| 35 | **The dead man's handle** | ALL · AGENTS | A train driver must keep a hand on the lever. Let go, and the train stops. The machine should stop when the human lets go, not carry on. | Railway Accident: Report on the Accident that occurred on 28th February 1975 at Moorgate Station (HMSO, 1976) – describes the driver's safety device and its limits. |
| 36 | **Go / no-go** | Q1 · M1 | Before lift-off, every station is asked in turn. One "no-go" holds the rocket. Nobody has to win the argument. One no is enough. | Gene Kranz, Failure Is Not an Option (2000) – the Mission Control go/no-go poll. |
| 37 | **No one judges his own case** | Q2 · M2 | The oldest rule of a fair hearing. A company testing its own product is judge and party at once. | Dr Bonham's Case (1610) 8 Co. Rep. 107a (Coke CJ) – nemo iudex in causa sua. |
| 38 | **Hammurabi's builder** | ALL | The one who builds carries the loss. One of the oldest written codes made the builder answer with his own life for a house that fell and killed its owner. AI carries no loss, so the say cannot be the machine's. | Code of Hammurabi (c. 1750 BC), law 229, trans. L. W. King (1910). |
| 39 | **The andon cord** | Q1 · M1 | Any worker can pull it and stop the whole factory. Oil rigs and nuclear plants call it stop-work authority. The no sits with the person who sees the fault, and nobody may punish them for it. | Taiichi Ohno, Toyota Production System: Beyond Large-Scale Production (1988); 30 CFR 250.1930 – stop work authority on US offshore installations; 10 CFR 50.7 – protection of nuclear workers who raise safety concerns. |
| 40 | **The external examiner** | Q2 | Marks are checked by a professor from another university. China did the older version a thousand years ago: exam papers sealed and copied out, so the examiner could not know whose work it was. | QAA, UK Quality Code for Higher Education – External Expertise (2018); John W. Chaffee, The Thorny Gates of Learning in Sung China (1985). |
| 41 | **Lagrådet (SE)** | Q2 | Before parliament votes on a law, senior judges say in public whether it holds. They cannot stop it, but their opinion is published with the bill, so a government that goes ahead anyway does so in the open. | Regeringsformen 8 kap. 20–22 §§; Lag (2003:333) om Lagrådet; the council was created in 1909. |

| # | Name | Tags | The idea | Reference |
|---|---|---|---|---|
| 42 | **Pre-registration** | Q3 · M3 | Before the trial starts, you write down what you are testing and what counts as a pass. Otherwise the pass mark moves to wherever the result landed. | De Angelis et al., "Clinical trial registration: a statement from the International Committee of Medical Journal Editors", New England Journal of Medicine 351 (2004), 1250–1251. |
| 43 | **Asilomar** | Q3 · M3 | When genetic engineering arrived, the scientists themselves stopped, wrote the safety rules, and then went on. The pause was not the point. The rules were. | Berg et al., "Potential Biohazards of Recombinant DNA Molecules", Science 185 (1974), 303; Berg et al., "Summary statement of the Asilomar Conference on Recombinant DNA Molecules", PNAS 72 (1975), 1981–1984. |
| 44 | **The co-pilot's no** | Q1 | Tenerife 1977, the deadliest accident in aviation history: a junior officer doubted the captain and did not insist. Since then the co-pilot's doubt is a rule the captain must answer. | Spanish accident report on the Tenerife collision (Comisión de Investigación de Accidentes, 1978); Cooper, White & Lauber (eds.), Resource Management on the Flightdeck, NASA CP-2120 (1980). |
| 45 | **One witness is no witness** | Q2 | A single account, however confident, proves nothing. One model checking one model is one witness. | Deuteronomy 19:15; Codex Justinianus 4.20.9 (Constantine, AD 334). |
| 46 | **The engineer's stamp** | Q1 · M1 | A licensed engineer signs the drawing in her own name, and if the bridge falls, it is her name. Canadian engineers wear an iron ring as a reminder; the ritual is popularly tied to a bridge that fell. | Professional Engineers Act, R.S.O. 1990, c. P.28 (Ontario) – the engineer's seal; The Ritual of the Calling of an Engineer (Rudyard Kipling, 1925). |
| 47 | **Independent verification** | Q2 · M2 | Software that can kill astronauts is checked by a separate team, under separate managers, from a separate budget. Independent means all three, because one is not enough. | IEEE Std 1012, Standard for System, Software, and Hardware Verification and Validation – technical, managerial and financial independence; NASA IV&V Program (established 1993). |
| 48 | **Countersignature, kontrasignation (SE)** | Q1 · M1 | The king's decision was void unless a minister also signed, and the minister, not the king, answered for it. An identifiable owner for every consequential decision, with the wider chain of evidence, challenge and authority preserved in the record. | 1809 års regeringsform, §§ 38 and 106–107. |
| 49 | **The Trial of the Pyx** | Q2 | Every year the Royal Mint's coins are weighed and tested – not by the Mint, but by a jury of goldsmiths, in public. Seven centuries on, the maker of money is still not trusted to check its own money. | Coinage Act 1971 (UK), s. 8; the trial is recorded from 1282. |
| 50 | **The unanimous verdict that acquits** | Q1 | If every judge voted guilty, the accused went free, because a court where nobody found one word to say for him had stopped looking. No dissent is a warning, not a comfort. | Babylonian Talmud, Sanhedrin 17a. |

*The examples are purely illustrative institutional analogies, not evidence that arrangements developed in other sectors can be transferred directly to AI. They were selected to illustrate recurring governance principles, including independent scrutiny, separation of functions, identifiable decision authority, reversibility and the capacity to halt consequential action. The analogies are necessarily imperfect: institutional arrangements differ across sectors, jurisdictions and periods, and their inclusion here does not imply that they have always operated successfully. Nor are they exhaustive. Their purpose is to broaden the range of institutional experience from which AI governance might learn, rather than to establish the effectiveness of any particular mechanism.*

## Appendix B: The independent body: suggested term sheet

Table 1 describes the body that Section 12 calls for, as a term sheet for discussion. It is the second rung, not the first. Nothing in Section 5 waits for it: the four requirements are public, and any reader can hold an answer to them today.

| | |
|---|---|
| **Purpose** | To receive the answers to the three questions, judge each against four requirements (a named person, a date, a number, and a stop exercised or independently tested, or an honest statement that neither has happened), and publish which are answers and which are not, with reasons. |
| **What it is not** | Not a regulator, not an evaluator of models, not a standards body. It reads answers about who holds control and says whether they hold up. It does not inspect weights and it does not certify safety. |
| **Relation to other bodies** | Complementary, not competing. The UN Independent International Scientific Panel on AI monitors the evidence on AI risk globally and has noted that no single organisation or country sees enough incidents to identify every pattern. This body does something narrower: it reads answers from named organisations about who holds control, and says whether they hold up. |
| **Precedents** | Lagrådet in Sweden: before parliament votes, senior judges say in public whether a proposed law holds. They cannot stop it, but a government that goes ahead has to explain itself in the open (Appendix A, item 41). The Trial of the Pyx in Britain: for more than seven centuries the Royal Mint's coins have been tested in public by outsiders, not by the Mint (item 49). This body has the same kind of power: not to stop, but to make going ahead without an answer visible. |
| **Form** | Independent foundation or association, registered in a jurisdiction with strong protection for published judgement (to be decided; Sweden, Switzerland and the Netherlands are candidates). Members are personally named. |
| **Seats** | Five to seven. One from AI safety research, one from public health or another field with mature accountability practice, one from institutional investment, one from law or supervision, one from a deploying sector. A chair chosen by the members. No seat for any laboratory or fund that is being judged. |
| **Funding** | A levy on participating laboratories and asset owners, paid into a pooled account administered at arm's length. No single contributor above 20 percent. Two years of funding committed before the first judgement is published. No funder may see a judgement before publication, other than as the organisation being judged. |
| **Independence test** | Who can remove a member, and who pays them. Members serve fixed three-year terms, removable only by a majority of the other members, and are paid from the pool. |
| **What it publishes** | The answers as received. Its judgement of each, with reasons. A running register of who has answered and who has not. The two numbers from each laboratory, tracked over time. An annual statement of its own funding, decisions and mistakes. Findings are descriptive: no ratings, no certificates. The organisation judged sees its judgement first, and its response is published beside it, unchanged. |
| **Cadence** | First judgements within eight weeks of the deadline. Updates every six months. Emergency statements when a stop is exercised or an incident is reported. |
| **Sunset** | The body dissolves if, after two years, no laboratory or fund has participated, and says so in public. It does not persist as a letterhead. |

*Table 1. Term sheet for the independent body, for discussion.*